\documentclass[pra,12pt,tightenlines]{revtex4}
\usepackage{latexsym}
\usepackage{graphicx}
\usepackage{amsmath}
\usepackage{amsbsy}
\usepackage{amsthm}
\usepackage{bbm}
\usepackage{bm}
\usepackage{dsfont} 

\begin{document}

\title{Relativistic entanglement of two massive particles}
\author{Nicolai Friis} \email{nicolai.friis@univie.ac.at}
\affiliation{Faculty of Physics, University of Vienna, Boltzmanngasse 5, A-1090 Vienna,
Austria}
\author{Reinhold A. Bertlmann} \email{reinhold.bertlmann@univie.ac.at}
\affiliation{Faculty of Physics, University of Vienna, Boltzmanngasse 5, A-1090 Vienna,
Austria}
\author{Marcus Huber} \email{marcus.huber@univie.ac.at}\affiliation{Faculty of Physics, University of Vienna, Boltzmanngasse 5, A-1090 Vienna,
Austria}
\author{Beatrix C. Hiesmayr} \email{beatrix.hiesmayr@univie.ac.at}
\affiliation{Faculty of Physics, University of Vienna, Boltzmanngasse 5, A-1090 Vienna,
Austria}
\affiliation{University of Sofia, Faculty of Physics, Blvd. James Bourchier 5, 1164 Sofia, Bulgaria}


\begin{abstract}

We describe the spin and momentum degrees of freedom of a system of two massive spin--$\tfrac{1}{2}$ particles as a 4 qubit system. Then we explicitly show how the entanglement changes between different partitions of the qubits, when considered by different inertial observers. Although the two particle entanglement corresponding to a partition into Alice's and Bob's subsystems is, as often stated in the literature, invariant under Lorentz boosts, the entanglement with respect to other partitions of the Hilbert space on the other hand, is not. It certainly does depend on the chosen inertial frame and on the initial state considered. The change of entanglement arises, because a Lorentz boost on the momenta of the particles causes a Wigner rotation of the spin, which in certain cases entangles the spin- with the momentum states. We systematically investigate the situation for different classes of initial spin states and different partitions of the 4 qubit space.

Furthermore, we study the behavior of Bell inequalities for different observers and demonstrate how the maximally possible degree of violation, using the Pauli-Lubanski spin observable, can be recovered by any inertial observer.

\end{abstract}

\maketitle

\section{Introduction}

For many years quantum entanglement has been challenging physicists not only with its abstract qualities, the puzzling examples of Erwin Schr\"odinger \cite{schroedinger35}, but also with its applications to deep physical problems such as the exclusion of local realistic theories via Bell's theorem and its variants \cite{bell-book,chsh69}. Recently, it has been of growing interest to study these problems in a relativistic setting \cite{czachor97}-\cite{moonetal04}. It was seen early in the discussion by Gingrich and Adami \cite{gingrichadami02} that the dependence of the Wigner-rotation on the momentum of the state causes difficulties to define spin entanglement in a Lorentz invariant manner, which also reflects in non-covariantly transforming reduced density matrices, see Ref.~\cite{peresetal02}. Proposing the entanglement of the total state, spin and momentum wave function, to be invariant, which we cannot conclude to be the case here, the problem seemed to be resolved.\\

In Ref.~\cite{alsingmilburn02} Alsing and Milburn then argued that the entanglement fidelity of the bipartite state is preserved under Lorentz transformations, whereas Jordan, Shaji, and Sudarshan \cite{jordanetal06b} found no sums of entanglements to be unchanged for certain types of two-particle states. In this work we use similar states as in \cite{jordanetal06b}, but in choosing specific parametrizations for the spin-states, we analyze in detail different partitions of the two-particle Hilbert space and the entanglement change of the overall state, occuring under Lorentz transformations. We can recover both the results of \cite{alsingmilburn02}, and \cite{jordanetal06b} as special cases of our calculations, but in contrast to the claim of \cite{jordanetal06b} we find that the sum of spin and momentum entanglement of the two-particle state is invariant for some states, e.g. if the spin state is given by the Bell state $\left|\,\psi^{\,+}\,\right\rangle$, while it changes for other spin states, e.g. for $\left|\,\psi^{\,-}\,\right\rangle$. For a large class of two-particle states we explicitely calculate the change of entanglement under Lorentz boosts for different partitions, finding analytic expressions, depending on the parameterization of the states, and value of the Wigner rotation angle.\\

Parallel to the debate about the Wigner-rotations, the discussions also centered around the question about the maximally possible violation of Bell inequalities in different inertial frames \cite{ahnetal02}-\cite{ahnetal03b}, with help of the spin observable used by Czachor \cite{czachor97} or variations thereof, see \cite{leechangyoung04}, as well as without, see \cite{chakrabarti09}. In Ref.~\cite{chakrabarti09} it is claimed that "entanglement is frame independent", which we take to mean that a maximally entangled state will only be completely disentangled for the limit of infinite momenta, i.e. when the Lorentz transformations go to the speed of light, which is why this statement is in agreement with our results. The second statement of \cite{chakrabarti09} "violation of Bell's inequality is frame-dependent" is only true if the measurement directions are not chosen appropriately, as we show in our analysis in Section \ref{sec:bellinequalities}, and as has been explicitly calculated by Lee and Chang-Young \cite{leechangyoung04}. Clearly, Bell's inequality need not be violated for an arbitrary choice of measurement directions in any frame but the maximal violation can be recovered for appropriate directions by any inertial observer.\\

The article is structured as follows, in Section \ref{sec:wigner rotations} we give a brief review of the Wigner-rotations, following mainly the reasoning of Ref.~\cite{alsingmilburn02}. In Section \ref{sec:massiveparticleentanglement} we analyze the effects of the Wigner-rotations on different classes of spin-momentum states and their entanglement by structuring the 2-particle states as a 4-qubit state of 2 spins and 2 momenta. The overall entanglement distributed over those 4-qubit states is then studied by simply calculating the mixedness of the several reduced density matrices. This entanglement monotone is then compared for the initial and the Lorentz boosted states, creating the interesting "entanglement egg-tray" of Figure \ref{fig:entanglementchangebelltype}. In Section \ref{sec:bellinequalities} we then present how the choice of the Pauli-Lubanksi spin observable provides a covariant meaning of spin expectation values, when Wigner-rotated systems are being considered and it is thus guaranteed to recover the maximal violation of a Bell inequality in all inertial frames.

\section{Wigner rotations of massive particles}\label{sec:wigner rotations}

In our discussion we denote the four spacetime coordinates with greek indices, e.g. $\mu, \nu$, running from 0 to 3, and four vectors by plain letters, while spatial vectors are denoted by arrows, e.g. $a=(a^{0},\vec{a})$, and latin indices run over the three spatial coordinates $\left\{1,2,3\right\}$. We choose the Minkowski metric to be $g_{\mu\nu}=\mathrm{diag}\left\{1,-1,-1,-1\right\}$, although this does not explicitly feature anywhere in our discussion. Furthermore we use units such that $\hbar=c=1$, so the four momentum of a massive particle is written as
\begin{equation}
p\,=\,\begin{pmatrix} p^{0} \\ \vec{p} \end{pmatrix}\ \ \ \mbox{and}\ \ \ p^{\mu}p_{\mu}\,=\,(p^{0})^{2}-\vec{p}^{\,2}\,=\,m^{2} \ \ \,.
\label{eq:massive particle four momentum}
\end{equation}
We write the single particle state of 4-momentum $p$ and spin $s$ -- we could include other properties as well, but those are not involved in our discussion, so the restriction to spin will suffice -- as
\begin{equation}
\left|\,p,\,s\,\right\rangle\,=\,\left|\,p\,\right\rangle\otimes\left|\,s\,\right\rangle \ \ \,,
\label{eq:single particle state}
\end{equation}
where sometimes subscripts labeling momentum and spin are added for clarity. We also want to emphasize here, that only momentum eigenstates, satisfying
\begin{equation}
P^{\,\mu}\,\left|\,p,\,s\,\right\rangle\,=\,p^{\,\mu}\,\left|\,p,\,s\,\right\rangle \ \ \ ,
\label{eq:momentum eigenstates}
\end{equation}
are used throughout this article, which we take to be normalized such that
\begin{equation}
\int\!d\mu(p)\,\left\langle\,p^{\,\prime}\!\right.\left|\,p\,\right\rangle\,=\,1 \ \ \,,
\label{eq:momentum eigenstate normalization}
\end{equation}
where $d\mu(p)$ is a suitable Lorentz-invariant integration measure. In a slight abuse of notation we dispose of the integration symbol in Eq.~(\ref{eq:momentum eigenstate normalization}), writing $\left\langle\,p^{\,\prime}\!\right.\left|\,p\,\right\rangle=1$ and imply an integration performed whenever scalar products involving momentum eigenstates are considered.

Considering now a homogeneous Lorentz transformation, denoted by $\Lambda$, we can represent it on the Hilbert space of our states as unitary operation $U(\Lambda)$, i.e.
\begin{equation}
\left|\,p_{\Lambda},\,s^{\,\prime}\,\right\rangle\,=\,U(\Lambda)\,\left|\,p,\,s\,\right\rangle\ \ \ .
\label{eq:unitary rep of lorentz trafo}
\end{equation}
Since we can write any 4-momentum as a Lorentz transformation acting on the rest frame momentum $k$ as $p\,=\,L(p)\,k$, we can rewrite equation (\ref{eq:unitary rep of lorentz trafo})
\begin{equation}
U(\Lambda)\,\left|\,p,\,s\,\right\rangle\,=\,U(\Lambda L(p))\,\left|\,k,\,s\,\right\rangle\,=\,
U(L(p_{\Lambda}))\,U(L^{-1}(p_{\Lambda})\Lambda L(p))\,\left|\,k,\,s\,\right\rangle \ \ \,,
\label{eq:unitary rep of lorentz trafo continued}
\end{equation}
where we have used the group structure of the Lorentz group and inserted the identity in the form $L(p_{\Lambda})L^{-1}(p_{\Lambda})$. Clearly, the second operator on the r.h.s of equation (\ref{eq:unitary rep of lorentz trafo continued}) is a rotation, since it takes the standard momentum $k$ first to $p$, then to $p_{\Lambda}=\Lambda p$ and back to $k$ again, resulting, at most, in a rotation, called the \emph{Wigner-rotation} $W(\Lambda ,p)$
\begin{equation}
\textrm{W}(\Lambda ,p)=L^{-1}(\mathrm{p_{\Lambda}})\Lambda L(\mathrm{p})\ \ \ ,
\label{eq:wigner rotation}
\end{equation}
such that we can express the transformed total state as
\begin{equation}
U(\Lambda)\,\left|\,p,\,s\,\right\rangle\,=\,
\sum\limits_{s^{\,\prime}}\,D_{s^{\,\prime}s}(W(\Lambda ,p))\,\left|\,p_{\Lambda},\,s\,\right\rangle\ \ \ .
\label{eq:wigner rotated total state}
\end{equation}

Here $D_{s^{\,\prime}s}(W(\Lambda ,p))$ is a suitable representation (corresponding to the respective spin of the particles) of Wigner's little group, whose elements leave the standard momentum $k$ invariant. In our case, spin--$\frac{1}{2}$ particles, we find the little group to be just $SU(2)\,$. Choosing the boosts $L(p)$ and $\Lambda$ such that particle and observer are moving in the $z$ and $x$ direction with velocities $\vec{u}$ and $\vec{v}$ respectively, we find the axis of rotation to be antiparallel to $\vec{u}\times\vec{v}$, while the angle $\delta$ of the rotation is given by

\begin{equation}
\tan\delta\,=\,\frac{\sinh\eta\,\sinh\xi}{\cosh\eta\,+\,\cosh\xi}\ \ \,,
\label{eq:wigner rotation angle}
\end{equation}
where $\eta$ and $\xi$ are the rapidities of the the boosts, given by $\tanh \eta=u$ and $\tanh \xi=v$. A more detailed discussion including also massless particles can be found in Ref.~\cite{alsingmilburn02}.

\section{Entanglement of two massive particles} \label{sec:massiveparticleentanglement}

Let us investigate the entanglement properties of two massive spin--$\tfrac{1}{2}$ particles under the effect of a Lorentz boost. We consider a demonstrative scheme for the motion of the particles and the observer to study the entanglement of the particles. We assume the particles to be in a state, where the spin and momentum degrees of freedom are initially separable from each other, i.e., the two particles in the unboosted frame are in a state of the form
\begin{equation}
\left|\,\psi\,\right\rangle_{\mathrm{total}}\,=\,\left|\,\psi\,\right\rangle_{\mathrm{mom}}\, \left|\,\psi\,\right\rangle_{\mathrm{spin}}\,.
	\label{eq:initial total state}
\end{equation}
The particles are taken to be moving along the $\pm z$ direction with equal amount, opposite directions, and sharp momenta, i.e. delta-distributions in momentum space, which is equivalent to demand that the momentum distributions be sufficiently narrow to result in single Wigner rotations. We want to assume, however, that the particle momenta can be in an entangled state. The resulting implications for the constructed total wavefunction regarding distinguishability and particle types will not concern us during the calculations but is briefly discussed in Section \ref{conclusion}. The momentum state $\left|\,\psi\,\right\rangle_{\mathrm{mom}}$ is chosen entangled and thus is parameterized by a single angle $\alpha$ as
\begin{equation}
\left|\,\psi\,\right\rangle_{\mathrm{mom}}\,=\,\cos\alpha\left|\,p_{+},p_{-}\,\right\rangle\,+
\,\sin\alpha\left|\,p_{-},p_{+}\,\right\rangle\ ,
	\label{eq:initial momentum state}
\end{equation}
where $p_{\pm}=(p^{0},\pm\,\vec{p}\,)^{T}$, which allows us to regard the momentum state in the qubit formalism, where $\left|\,p_{+},p_{-}\,\right\rangle$ can be treated analogously to the qubit state $\left|\,1,0\,\right\rangle$. With this choice of the momentum state (\ref{eq:initial momentum state}) the initial state (\ref{eq:initial total state}) is transformed into the boosted state $\left|\,\psi^{\,\Lambda}\,\right\rangle_{\mathrm{total}}$, which exhibits entanglement between the spin- and momentum degrees of freedom, of course, depending on the value of $\alpha$ and the explicit action of the Wigner rotations on the spin state $\left|\,\psi\,\right\rangle_{\mathrm{spin}}$. It is of the form
\begin{eqnarray}
	\left|\,\psi^{\,\Lambda}\,\right\rangle_{\mathrm{total}} &=&
		\,\cos\alpha\left|\,\Lambda p_{+},\Lambda p_{-}\,\right\rangle\,
		\left(U_{+}\otimes\,U_{-}\right)\,\left|\,\psi\,\right\rangle_{\mathrm{spin}} \nonumber \\
	&+& \,\sin\alpha\left|\,\Lambda p_{-},\Lambda p_{+}\,\right\rangle\,
		\left(U_{-}\otimes\,U_{+}\right)\,\left|\,\psi\,\right\rangle_{\mathrm{spin}}\ ,
	\label{eq:boosted general total state}
\end{eqnarray}
where $U_{\pm}$ are the Wigner rotations about a yet to be fixed axis corresponding to the momenta $p_{\pm}$ and $\Lambda$ represents the Lorentz transformation. Clearly, if neither $\sin\alpha$ nor $\cos\alpha$ vanish, and the rotated spin states are unequal, the total state will not factorize and thus will be entangled between spin and momentum. Since we did not specify the spin state yet, we cannot claim that the overall entanglement of the state is changed. This, however, is to be expected, since the operation performed on the spin state cannot be written as one tensor product of unitary operations on the individual spin Hilbert spaces, though each operation is in itself unitary. Thus it does not qualify as being termed a "local unitary operation" but it can be viewed as a kind of double C-NOT gate, where the two control qubits and the two input qubits are allowed to be in entangled states respectively.\\

Let us now continue by examining different classes of initial spin states $\left|\,\psi\,\right\rangle_{\mathrm{spin}}$, namely coherent superpositions of the Bell states $\left|\,\psi^{\,\pm}\,\right\rangle$ at first and coherent superpositions of the spin triplet states later on, where  the z-axis is always choosen as the axis of spin quantization.

\subsection{Bell $\psi^{\,\pm}$ spin states}\label{subsec:bell type states}

Starting with the states $\left|\,\psi^{\,\pm}\,\right\rangle = 1/\sqrt{2}\,\left(\left|\,\uparrow\,\downarrow\,\right\rangle\,\pm\,\left|\,\downarrow\,\uparrow\,\right\rangle\right)$, we utilize a similar parametrization as for the momentum state (\ref{eq:initial momentum state}) earlier
\begin{equation}
\left|\,\psi\,\right\rangle_{\mathrm{spin}}\,=\,\cos\beta\left|\,\uparrow\,\downarrow\,\right\rangle\,+\,
\sin\beta\left|\,\downarrow\,\uparrow\,\right\rangle\ .
	\label{eq:initial spin state bell type}
\end{equation}
So the unboosted observer describes the total system by the state
\begin{equation}
\left|\,\psi\,\right\rangle_{\mathrm{total}}\,=\,\left(\,\cos\alpha\left|\,p_{+},p_{-}\,\right\rangle\,+\,
\sin\alpha\left|\,p_{-},p_{+}\,\right\rangle\ \right)
\left(\,\cos\beta\left|\,\uparrow\,\downarrow\,\right\rangle\,+\,
\sin\beta\left|\,\downarrow\,\uparrow\,\right\rangle\ \right)\,.
	\label{eq:initial total state bell type spin}
\end{equation}

Since the total state (\ref{eq:initial total state}), represented by the density operator
\begin{equation}
\rho\,=\,\left|\,\psi\,\right\rangle\left\langle\,\psi\,\right| \ \ ,
\label{eq:initial state Bell type spin density matrix}
\end{equation}
where $\left|\,\psi\,\right\rangle$ is taken to be the state (\ref{eq:initial total state bell type spin}), is a pure state, we can calculate the amount of entanglement, distributed between the different partitions of the 4 qubits (2 spin- and 2 momentum qubits), by calculating and adding the linear entropies of the corresponding reduced density matrices (see e.g. Ref.~\cite{mintertcarvalhokusbuchleitner05}), i.e.
\begin{equation}
E(\rho)\,=\,\sum\limits_{i}\,\left(\,1-\rm{Tr}\,\rho_{i}^{\,2}\,\right) \ ,
 \label{eq:linear entropy}
\end{equation}
where $\rho_{i}$ is obtained by tracing over all subsystems except the $i$-th.

We consider formula ({\ref{eq:linear entropy}) as an appropriate entanglement measure, since it reaches a maximal value of 1 for a maximally entangled 2-qubit state but other conventions can be easily obtained by a linear rescaling, typically by a factor of 2. We use this measure now to calculate the entanglement between the different possible partitions of the 4 qubits.\\

Let's begin with investigating the entanglement of 1 qubit in relation to the other 3 remaining qubits, thus one subsystem contains 1 qubit the other 3 qubits. Then we get for the state (\ref{eq:initial total state bell type spin}) the total amount of entanglement
\begin{equation}
E(\rho)\;=\;\tfrac{1}{2}\left(\,2\,-\,\cos(4\alpha)\,-\,\cos(4\beta)\,\right) \,.
\label{eq:linear entropy of unboosted bell type state}
\end{equation}
Note, that for $\alpha=\beta=\tfrac{(2n+1)\pi}{4}$, i.e., both spin and momentum in the Bell states $\left|\,\psi^{\,\pm}\,\right\rangle$, we have maximal entanglement, whereas for $\alpha=\beta=\tfrac{n\pi}{2}$, i.e. a fully separable state, the linear entropy vanishes.\\

Assume now, that there is a second observer moving in the $x$-direction, such that her or his frame is related to our system by a boost along the $(-x)$-direction, which, due to the particle momenta in the $(\pm z)$-direction, will result in Wigner rotations around the $(-y)$-direction about angles $\pm \delta$ respectively. This means, that our Wigner rotation matrices $U_{\pm}$ from Eq.~(\ref{eq:boosted general total state}) will be of the form
\begin{equation}
U_{\pm}\,=\,\begin{pmatrix} \cos\tfrac{\delta}{2} & \pm\sin\tfrac{\delta}{2} \\
														\mp\sin\tfrac{\delta}{2} & \cos\tfrac{\delta}{2}\end{pmatrix}\ .
\label{eq:rotation matrices}
\end{equation}

The calculation of the boosted state $\left|\,\psi^{\,\Lambda}\,\right\rangle_{\mathrm{total}}$ according to equation (\ref{eq:boosted general total state}) and the corresponding density matrix $\rho^{\,\Lambda}$ is straightforward
\begin{eqnarray}
	\left|\,\psi^{\,\Lambda}\,\right\rangle_{\mathrm{total}} &=&
		\,\cos\alpha\left|\,\Lambda p_{+},\Lambda p_{-}\,\right\rangle\,
		\left[\,c_{1}\,\left(\left|\,\uparrow\uparrow\,\right\rangle\,+
		\,\left|\,\downarrow\downarrow\,\right\rangle\,\right)\,+
		\,c_{2}\,\left|\,\uparrow\downarrow\,\right\rangle\,+
		\,c_{3}\,\left|\,\downarrow\uparrow\,\right\rangle\,\right]+ \nonumber \\
	&+& \,\sin\alpha\left|\,\Lambda p_{-},\Lambda p_{+}\,\right\rangle\,
		\left[\,-c_{1}\,\left(\left|\,\uparrow\uparrow\,\right\rangle\,+
		\,\left|\,\downarrow\downarrow\,\right\rangle\,\right)\,+
		\,c_{2}\,\left|\,\uparrow\downarrow\,\right\rangle\,+
		\,c_{3}\,\left|\,\downarrow\uparrow\,\right\rangle\,\right]	\,,
\label{eq:boosted bell type spin total state}
\end{eqnarray}
where $c_{1}=\tfrac{1}{2}\sin\delta\,(\sin\beta-\cos\beta)$, $c_{2}=\cos\beta\,\cos^{2}\!\tfrac{\delta}{2}+\sin\beta\,\sin^{2}\!\tfrac{\delta}{2}$, $c_{3}=\sin\beta\,\cos^{2}\!\tfrac{\delta}{2}+\cos\beta\,\sin^{2}\!\tfrac{\delta}{2}$ and
\begin{equation}
\rho^{\,\Lambda}\,=\,\left|\,\psi^{\,\Lambda}\,\right\rangle\left\langle\,\psi^{\,\Lambda}\,\right|\,.
\label{eq:boosted bell type spin total density matrix}
\end{equation}

Calculating the linear entropy of this state we find the result
\begin{eqnarray}
E(\rho^{\,\Lambda}) &\;=\;& \tfrac{1}{16}\left(\right.18\,-\,10\cos(4\alpha)\,-\,6\cos(4\beta) 		
		\,-\,2\cos(4\alpha)\cos(4\beta)	\nonumber \\
 && -\,8\cos(2\delta)\sin^{2}(2\alpha)\cos^{2}(2\beta)\left.\right)\,,
\label{eq:linear entropy of boosted bell type state}
\end{eqnarray}
which approaches expression (\ref{eq:linear entropy of unboosted bell type state}) for vanishing Wigner angle $\delta \rightarrow 0\,$.
The difference of the linear entropies in the boosted and unboosted system emerges then quite simple
\begin{equation}
E(\rho^{\,\Lambda})-E(\rho)\;=\;\sin^{2}\!\delta\,\sin^{2}(2\alpha)\,\cos^{2}(2\beta)\,.
\label{eq:linear entropy difference of boosted and unboosted bell type state}
\end{equation}

This expression is easy to analyze. We immediately see that for $\delta=0$, i.e. no change of the inertial frame, the difference vanishes as expected. The interesting fact about Eq.~(\ref{eq:linear entropy difference of boosted and unboosted bell type state}) is, however, that the overall entanglement of the state \textit{does} change, depending on the choice of initial state, as well as the strength of the boost. If for instance $\alpha$ is chosen to be $\tfrac{n\pi}{2}$, where $n$ can be any integer, the initial momentum state becomes separable and the entanglement does not change after the boost. Considering, on the other hand, initially some entanglement in the momentum part, then increasing the entanglement in the momentum state implies also an increase in the difference of the linear entropies, creating the striking "egg-tray" pattern which can be seen in Fig.~\ref{fig:entanglementchangebelltype}.

\begin{figure}[ht]
	\centering
		\includegraphics[width=0.80\textwidth]{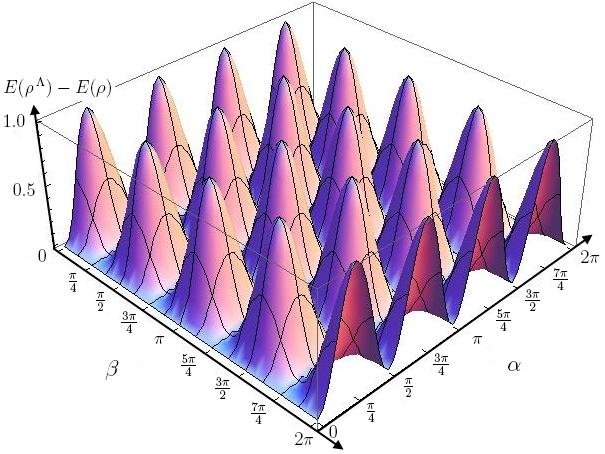}
	\caption{Entanglement-Egg-Tray: Difference of linear entropy of a $\delta =\pm\tfrac{\pi}{2}$ Wigner rotated Bell-type two particle state in case of partition 1 qubit versus 3 qubits. Plot of Eq.~(\ref{eq:linear entropy difference of boosted and unboosted bell type state}).}
	\label{fig:entanglementchangebelltype}
\end{figure}

We also notice that for the maximally entangled Bell states $\left|\,\psi^{\,\pm}\,\right\rangle\,$, corresponding to a choice of $\beta=\tfrac{(2n+1)\pi}{4}$, the entanglement does not change regardless of the Wigner rotation angle or the choice of $\alpha$ in the momentum state. This includes, e.g., the example given by Ref.~\cite{chakrabarti09}, where the momentum state is separable and the spin state is totally antisymmetric. If, however, the entanglement in the spin decreases we obtain again an increase in the entropy difference (see Fig.~\ref{fig:entanglementchangebelltype}).\\

Next we choose another partition of the 4 qubit Hilbert space, we consider 2 qubits in each subsystem and start with investigating the entanglement of the 2 spin qubits in relation to the 2 momentum qubits. We proceed in an analogue manner as before, i.e. starting from Eqs.~(\ref{eq:initial total state bell type spin}) and (\ref{eq:initial state Bell type spin density matrix}) we calculate the reduced density matrices for the two spin- or momentum-degrees of freedom
\begin{equation}
\rho_{\mathrm{spin}}\,=\,\mathrm{Tr}_{\mathrm{mom}}(\rho)\ ,\ \ \ \rho_{\mathrm{mom}}\,=\,\mathrm{Tr}_{\mathrm{spin}}(\rho)
\label{eq:reduced density matrices for spin and momentum}
\end{equation}
and their respective mixednesses. Clearly, the linear entropy (\ref{eq:linear entropy}) of the total state is identically zero, $E(\rho)=0$, since the initial state (\ref{eq:initial total state bell type spin}) factorizes with respect to this partition. Repeating this process for the boosted state (\ref{eq:boosted bell type spin total state}) we find, however, that the entanglement w.r.t. this partition does not vanish
\begin{equation}
E(\rho^{\,\Lambda})\;=\;\frac{1}{2}\,
\sin^{2}\!\delta\,\sin^{2}(2\alpha)\,\left(1\,-\,\sin(2\beta)\right)\,\left[3\,+\,\cos(2\delta)\,+\,
2\sin^{2}\!\delta\sin(2\beta)\right] \ \ \,.
\label{eq:spin vs momentum entanglement boosted bell type state}
\end{equation}

\begin{figure}[h]
	\centering
		\includegraphics[width=0.70\textwidth]{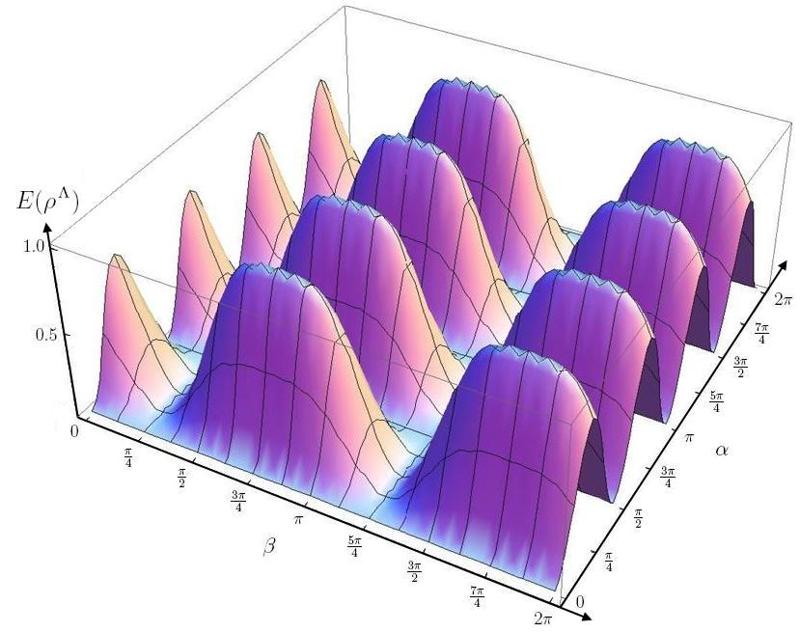}
	\caption{Entanglement in case of partition 2 spins versus 2 momenta of a $\delta =\pm\tfrac{\pi}{4}$ Wigner rotated spin-Bell-type state. Plot of Eq.~(\ref{eq:spin vs momentum entanglement boosted bell type state}).}
	\label{fig:spinmomentumentanglementbelltype}
\end{figure}

Clearly, formula (\ref{eq:spin vs momentum entanglement boosted bell type state}) presents a similar dependence on the Wigner rotation angle $\delta$ and the initial entanglement of the momentum state, parameterized by $\alpha$ as the equation (\ref{eq:linear entropy difference of boosted and unboosted bell type state}). But as can be seen in Fig.~\ref{fig:spinmomentumentanglementbelltype}, although the valleys of the plot agree for $\alpha=\frac{n\pi}{2}$ the entanglement change due to the boost is no longer zero for all values $\beta=\frac{(2n+1)\pi}{4}$ as in Fig.~\ref{fig:entanglementchangebelltype}. This points to an imbalance between the Bell states $|\psi^{\,-}\rangle$ and $|\psi^{\,+}\rangle$ since the overall entanglement does not change for either of the two. While the entanglement distributed between momentum and spin is only invariant for $\beta=\frac{(4n+1)\pi}{4}\,$, i.e. for the symmetric state $|\psi^{\,+}\rangle\,$, it does change for $\beta=\frac{(4n+3)\pi}{4}\,$, i.e. the antisymmetric state $|\psi^{\,-}\rangle\,$, recovering such the results of Ref.~\cite{jordanetal06b}.

Interestingly, if we consider the limit $\delta\rightarrow\tfrac{\pi}{2}\,$, corresponding to observer and particle moving with velocity of light, expression (\ref{eq:spin vs momentum entanglement boosted bell type state}) coincides with the simple formula (\ref{eq:linear entropy difference of boosted and unboosted bell type state}) and we recover the entanglement-egg-tray of Fig.~\ref{fig:entanglementchangebelltype}. Thus the entanglement change of the above discussed two partitions of qubits (leading to (\ref{eq:linear entropy difference of boosted and unboosted bell type state}) and (\ref{eq:spin vs momentum entanglement boosted bell type state})) is the same in this limit.\\

Finally, we come to the important case of the entanglement between the two moving particles with spin--$\tfrac{1}{2}$, i.e. we consider the partition into the Hilbert spaces containing the momentum qubit and the spin qubit of each particle, which we want to call the \emph{Alice-Bob partition}. Here the reduced density matrices for Alice and Bob are obtained by tracing over the momentum and spin of the complementary subspace
\begin{equation}
\rho^{\,A}_{\mathrm{mom-spin}}\,=\,\mathrm{Tr}^{\,B}_{\mathrm{mom-spin}}(\rho)\ ,\ \ \ \rho^{\,B}_{\mathrm{mom-spin}}\,=\,\mathrm{Tr}^{\,A}_{\mathrm{mom-spin}}(\rho) \ \ \,.
\label{eq:reduced density matrices for alice and bob}
\end{equation}
Calculating the linear entropy we find the following expression
\begin{equation}
E(\rho)\;=\;\tfrac{1}{8}\left[\,10\,-\,\left(3+\cos(4\alpha)\right)\,\left(3+\cos(4\beta)\right)\,\right] \,,
\label{eq:linear entropy of unboosted bell type spin-mom state}
\end{equation}
which clearly vanishes for $\alpha=\beta=\tfrac{n\pi}{2}$, i.e. a fully separable state. Whereas for $\alpha=\beta=\tfrac{(2n+1)\pi}{4}$, i.e. both spin and momentum in the Bell states, we have maximal entanglement $E=\frac{3}{2}$, which means the reduced density matrices for Alice and Bob are maximally mixed $\rho^{\,A}_{\mathrm{mix}}=\rho^{\,B}_{\mathrm{mix}}=\frac{1}{4}\mathbbm{1}_4\,$ and $\,1-\mathrm{Tr}(\rho^A)^2=\frac{3}{4}\,$.

Performing now a boost of the system along the $(-x)$-direction we interestingly find that the entanglement with respect to the Alice-Bob partition does \emph{not} change at all, i.e. $E(\rho^{\,\Lambda})=E(\rho)$, regardless of the parametrization of the state or the strength of the boost. This result is quite in accordance with the maintained violation of a Bell-inequality (see Section~\ref{sec:bellinequalities}), sensitive to exactly this partition of the Hilbert space.

Remarkably furthermore, by tracing over spin and momentum it does not matter which particle the spin and momentum belongs to. We may trace over the spin of particle 1 and momentum of particle 2 (or vice versa) and obtain the same result (\ref{eq:linear entropy of unboosted bell type spin-mom state}), there is \emph{no} change in the entanglement for all possible states after the boost. This may reflect the nonlocal feature of quantum mechanics.

\subsection{Spin triplet states}\label{subsec:triplet type states}

To further investigate the properties of two particle spin states under Lorentz transformations we now consider coherent superpositions of spin triplet states. The singlet state is already included in our previous discussion. We now choose spherical coordinates in the three dimensional space of the triplet states to parameterize all possible combinations. The spin state in Eqs. (\ref{eq:initial total state}), (\ref{eq:boosted general total state}) is thus replaced by
\begin{equation}
\left|\,\psi\,\right\rangle_{\mathrm{spin}}\,=\,\sin\theta\,\cos\phi\left|\,\uparrow\,\uparrow\,\right\rangle\,+\,
\sin\theta\,\sin\phi\,\frac{1}{\sqrt{2}}\left(\left|\,\uparrow\,\downarrow\,\right\rangle\,+\,
\left|\,\downarrow\,\uparrow\,\right\rangle\right)\,+\,
\cos\theta\,\left|\,\downarrow\,\downarrow\,\right\rangle\ .
	\label{eq:initial spin state triplet type}
\end{equation}
Using the same momentum state and inertial frames, and studying the same partitions of the qubits as in Subsection \ref{subsec:bell type states}, we once again compute the difference of the linear entropy of the boosted and unboosted observer.\\

Beginning with the partition 1 qubit versus 3 qubits, the difference of the entanglement turns out to be
\begin{eqnarray}
E(\rho^{\,\Lambda})-E(\rho) &\;=\;&  		
	-\frac{1}{4}\sin^{2}\!\delta\sin^{2}(2\alpha)\,\left(\cos\theta+\cos\phi\sin\theta\right)^{2}
	\left[\,-\,5\,+\right.\nonumber \\
&& +\left.\cos(2\theta)+2\sin^{2}\!\theta\,\cos(2\phi)+4\sin(2\theta)\,\cos\phi\right]\,.
\label{eq:linear entropy difference of boosted and unboosted triplet type state}
\end{eqnarray}

\begin{figure}
	\centering
		\includegraphics[width=0.80\textwidth]{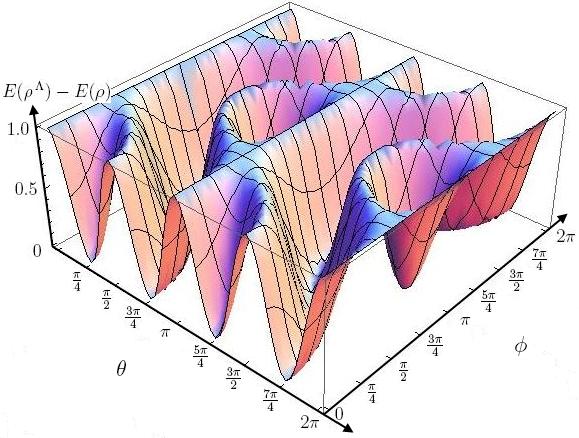}
	\caption{Difference of linear entropy of a $\delta =\pm\tfrac{\pi}{2}$ Wigner rotated spin-triplet-type state with totally symmetric momentum state $(\alpha=\tfrac{\pi}{4})$ in case of partition 1 qubit versus 3 qubits. Plot of Eq.~(\ref{eq:linear entropy difference of boosted and unboosted triplet type state}).}
	\label{fig:entanglementchangetripletttype}
\end{figure}

In formula (\ref{eq:linear entropy difference of boosted and unboosted triplet type state}) the spin parametrization (previously represented by $\beta$) appears now a bit more involved, but this is only due to the choice of the parametrization of the spherical coordinates $(\theta,\phi)\,$, the dependence on the Wigner rotation angle $\delta$ and the momentum parameter $\alpha$ is exactly the same as before (compare with Eq.~(\ref{eq:linear entropy difference of boosted and unboosted bell type state})). Interestingly, the change of entanglement comes out bigger the less entangled the initial spin state is. On the other hand, maximally entangled spin states such as the Bell states $\left|\,\psi^{\,\pm}\right\rangle$, $\left|\,\Phi^{\,\pm}\right\rangle = 1/\sqrt{2}\,\left(\left|\,\uparrow\,\uparrow\,\right\rangle\,\pm\,\left|\,\downarrow\,\downarrow\,\right\rangle\right)$ or certain superpositions such as $\frac{1}{N}\left(a\left|\,\Phi^{\,\mp}\right\rangle+b\left|\,\psi^{\,\pm}\right\rangle\right)$ with $N^2=|a|^2+|b|^2$, which are also maximally entangled, lie all on the bottom of the valleys of Fig.~\ref{fig:entanglementchangetripletttype}. That means, the entanglement of the total states does not change under Lorentz transformations, even though the form of the states does, or put differently, the total entanglement is saturated for maximally entangled spin states and therefore cannot be further increased.\\

To give an example, consider the initial state
\begin{equation}
\left(\,\cos\alpha\left|\,p_{+},p_{-}\,\right\rangle\,+\,\sin\alpha\left|\,p_{-},p_{+}\,\right\rangle\right)\,
\left|\,\Phi^{\,+}\right\rangle
\label{eq:example initial spin state phi plus}
\end{equation}
corresponding to a choice of $\phi=0$ and $\theta=\tfrac{\pi}{4}$, which in our setup is transformed to
\begin{eqnarray}
\longrightarrow\ \ && \cos\alpha\left|\,\Lambda p_{+},\Lambda p_{-}\,\right\rangle
		\left(\,\cos\delta\,\left|\,\Phi^{\,+}\right\rangle\,+
		\,\sin\delta\,\left|\,\psi^{\,-}\right\rangle\,\right) \,+\nonumber \\
&+& \,\sin\alpha\left|\,\Lambda p_{-},\Lambda p_{+}\,\right\rangle
		\left(\,\cos\delta\,\left|\,\Phi^{\,+}\right\rangle\,-
		\,\sin\delta\,\left|\,\psi^{\,-}\right\rangle\,\right)\ .
\label{eq:example transformed spin state phi plus}
\end{eqnarray}
It has the same linear entropy, $E(\rho)=\tfrac{1}{2}(3-\cos(4\alpha))$, as the initial state.

Let us now consider an example of an initially separable spin state, of both spins orientated in the $+z$ direction, i.e. $\phi=0$ and $\theta=\tfrac{\pi}{2}$,
\begin{equation}
\left(\,\cos\alpha\left|\,p_{+},p_{-}\,\right\rangle\,+\,\sin\alpha\left|\,p_{-},p_{+}\,\right\rangle\right)\,
\left|\,\uparrow\uparrow\,\right\rangle\,,
\label{eq:example initial spin state upup}
\end{equation}
it becomes an overall entangled state for a suitable choice of parameters $\alpha$ and $\delta$. The difference in the entanglement, the difference of the linear entropies (\ref{eq:linear entropy difference of boosted and unboosted triplet type state}), before and after the boost results in
\begin{equation}
E(\rho^{\,\Lambda})-E(\rho)\;=\;\sin^{2}(2\alpha)\sin^{2}\!\delta\,.
\label{eq:example initial spin state upup entanglement difference}
\end{equation}
This entropy change (\ref{eq:example initial spin state upup entanglement difference}) becomes maximal, if the initial momentum state is maximally entangled, $\alpha=\tfrac{\pi}{4}$, and the speed of the boosted observer and the particles approach the speed of light, $\delta\rightarrow\tfrac{\pi}{2}$.\\

Considering next the partition into 2 spin qubits versus 2 momentum qubits, i.e. tracing over the momenta or spins, we find for the entanglement change ($E(\rho)=0$ in this partition)
\begin{eqnarray}
E(\rho^{\,\Lambda}) &\;=\;& 1-\cos^{4}\!\alpha-\sin^{4}\!\alpha 		
	\,+\,\frac{1}{32}\sin^{2}\!\delta\sin^{2}(2\alpha)\,\left(\cos\theta+\cos\phi\sin\theta\right)^{2}
	\left(\,26+f_1-f_2\,\right)\nonumber \\
&& -\frac{1}{512}\sin^{2}(2\alpha)\,\left(\,10+f_1-f_2\,\right)^2 \ \ \,,
\label{eq:spin versus momentum entanglement of boosted triplet type state}
\end{eqnarray}
where the functions $f_1,f_2$ are defined by
\begin{eqnarray}
f_1 &\;=\;&f_1(\delta,\theta)\;=\;2\cos(2\delta)\,\left(\,3+\cos(2\theta)\,\right)\,-\,2\cos(2\theta)\\
f_2 &\;=\;&f_2(\delta,\theta,\phi)\;=\;8\sin^{2}\!\delta\,\left(\,\cos(2\phi)\sin^2\!\theta\,+\,
2\cos\phi\sin(2\theta)\,\right) \ \ \,.
\label{eq:functions f1 and f2 for spin versus momentum entanglement of boosted triplet type state}
\end{eqnarray}

\begin{figure}
	\centering
		\includegraphics[width=0.80\textwidth]{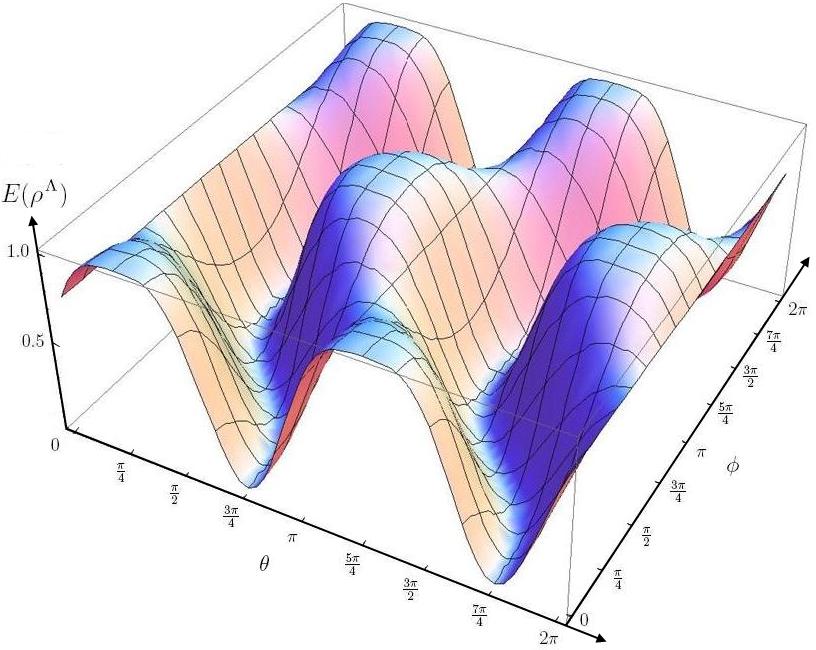}
	\caption{Entanglement in case of partition 2 spins versus 2 momenta of a $\delta =\pm\tfrac{\pi}{4}$ Wigner rotated spin-triplet-type state with totally symmetric momentum state $(\alpha=\tfrac{\pi}{4})$. Plot of Eq.~(\ref{eq:spin versus momentum entanglement of boosted triplet type state}).}
	\label{fig:spinmomentumentanglementchangetripletttype}
\end{figure}

Clearly, the boosted result (\ref{eq:spin versus momentum entanglement of boosted triplet type state}) is more involved than the one (\ref{eq:linear entropy difference of boosted and unboosted triplet type state}) of the previous partition but, as we can see from Fig.~\ref{fig:spinmomentumentanglementchangetripletttype}, the valleys starting at $\theta=\frac{3\pi}{4},\frac{7\pi}{4}, ...\,$ and $\phi=0$ remain. However, there is an entanglement change between the valleys, which is different to the previously considered partition (see Fig.~\ref{fig:entanglementchangetripletttype}). As already mentioned, it points to an imbalance between the Bell states, in this case $|\phi^{\,+}\rangle$ and $|\psi^{\,+}\rangle\,$, $|\phi^{\,-}\rangle$. Considering furthermore the limit of speed of light, $\delta\rightarrow\tfrac{\pi}{2}\,$, then the expressions (\ref{eq:spin versus momentum entanglement of boosted triplet type state}) and (\ref{eq:linear entropy difference of boosted and unboosted triplet type state}) agree (and thus Fig.~\ref{fig:spinmomentumentanglementchangetripletttype} and Fig.~\ref{fig:entanglementchangetripletttype}), as we already could anticipate from our study of the Bell-type states.\\

Studying finally the case of the Alice-Bob partition, i.e. tracing over spin and momentum of a subspace, the linear entropy turns out to be
\begin{eqnarray}
E(\rho) &\;=\;& \frac{1}{256}\Big[\,203\,-\,103\cos(4\alpha)\,+\,
\left(\,3+\cos(4\alpha)\,\right)\ \nonumber \\
&&\,\times\,\big(\,-12\cos(2\theta)\,-\,13\cos(4\theta)\,+\,
16\left(\,3+5\cos(2\theta)\,\right)\cos(2\phi)\sin^2\!\theta \nonumber \\
&&\,+\,8\cos(4\phi)\sin^4\!\theta\,-\,256\cos\theta\cos\phi\sin^3\!\theta\sin^2\!\phi\big)\Big] \ \ \,.
\label{eq:linear entropy of Alice-Bob partition of unboosted triplet type state}
\end{eqnarray}
As expected, the linear entropy vanishes for $\alpha=\tfrac{n\pi}{2}$ and $\theta=n\pi$,  i.e. a fully separable state, whereas for $\alpha=\tfrac{(2n+1)\pi}{4}$ and  either $\theta=\phi=\tfrac{(2n+1)\pi}{2}$, or $\theta=\tfrac{(2n+1)\pi}{4},\, \phi=n\pi$, i.e. both momentum and spin in the Bell states, we have maximal entanglement $E=\frac{3}{2}$.

Performing the boost of the system along the $(-x)$-direction we find, as in the case of the Bell spin states, that the entanglement with respect to the Alice-Bob partition does \emph{not} change at all, i.e. $E(\rho^{\,\Lambda})=E(\rho)$, regardless of the parametrization of the state or the strength of the boost.
As found before, we also may trace over the spin of particle 1 and momentum of particle 2 (or vice versa) and the entanglement does not change at all for all possible states after the boost.

\section{Violation of Bell inequalities}\label{sec:bellinequalities}

Since we have established the fact that the overall entanglement of a state, composed in the manner of (\ref{eq:initial spin state bell type}) or (\ref{eq:initial spin state triplet type}), is generally not the same for all inertial observers, we now want to analyze how this affects physical applications of entanglement, such as the violation of a Bell inequality. A typical setup for such an (gedanken-)experiment with massive particles could consist of two spin--$\tfrac{1}{2}$ particle beams emitted by some source in opposite directions along the $(\pm z)$-direction, into which they propagate until they interact with measurement devices, e.g. two Stern-Gerlach apparatuses, which we call Alice (A) and Bob (B) respectively. The described setup now also implies a certain choice of reference frame, namely one in which the particle momenta are confined to two possible values $\pm\vec{p}$ (ignoring spreads in momentum space) along the $z$-axis and the detectors A and B are at rest w.r.t. the source. Therefore first we need to find the correct observable for spin measurements in different reference frames.\\

Previous analysis \cite{czachor97,ahnetal02,ahnetal03a,moonetal04} suggested to use a spin observable $\vec{\sigma}_{p}\,=\,\overrightarrow{W}/p^{0}$, closely related to the Pauli-Lubanski vector $W_{\mu}$ \cite{ryderqft}, which is defined by $W_{\mu}=-\frac{1}{2}\varepsilon_{\mu\nu\varrho\sigma}J^{\nu\varrho}p^{\sigma}$ and $J_{\nu\varrho}=\{J_{ij}=\varepsilon_{ijk} J_k \;\; / \;\; J_{0j}=K_j \;\;\mbox{with}\;\; K_j=i\,(t\,\partial_j+x_j\partial_t)\}$ contains the total angular momentum or spin $J_k$. Its squared $W_{\mu}W^{\mu}$ is one of the Casimir operators of the Poincar$\acute{e}$ group, i.e. a conserved quantity. The corresponding normalized binary observable for a spin measurement along direction $\vec{a}$ in a frame where the particle has momentum $p=(p^{0},\vec{p}\,)$ is given by
\begin{equation}
\hat{a}(p)\,=\,\frac{\vec{a}\cdot\vec{\sigma}_{p}}{|\lambda(\vec{a}\cdot\vec{\sigma}_{p})|} \ \ \,,
\label{eq:normalized relativistic spin observable}
\end{equation}
where $\lambda(\vec{a}\cdot\vec{\sigma}_{p})$ is the eigenvalue of the operator $\vec{a}\cdot\vec{\sigma}_{p}$. It can be re-expressed by
\begin{equation}
\hat{a}(p)\,=\,\vec{a}_{p}\cdot\vec{\sigma} \quad \; \mbox{with} \quad \; \vec{a}_{p}\,=\,\frac{\sqrt{1-\beta^{2}}\,\vec{a}_{\perp}\,+\,\vec{a}_{\parallel}}
{\sqrt{1\,+\,\beta^{2}(\vec{a}_{\parallel}^{\,2}\,-\,1)}} \ \ \,,
\label{eq:pauli-lubanski-direction}
\end{equation}
with $\beta$ being the velocity (we have set $c=1$) of the particle in the frame and
$\vec{\sigma}$ is the usual vector of Pauli-matrices. The vector $\vec{a}_{p}$ is a unit vector and can be interpreted as the detector orientation $\vec{a}$ as seen from the particle rest frame, i.e. we can rewrite equation (\ref{eq:pauli-lubanski-direction}) as
\begin{equation}
\hat{a}(p)\,=\,\vec{a}_{p}\cdot\vec{\sigma}\,=\,\frac{(L^{-1}(p)a)^{i}\sigma_{i}}{|(L^{-1}(p)a)^{j}|} \ \ \,,
\label{eq:pauli-lubanski-direction from rest frame}
\end{equation}
where $|(L^{-1}(p)a)^{j}|$ is the norm of the spatial part of the Lorentz transformed orientation vector $L^{-1}(p)a\,$.

As has been demonstrated by Czachor \cite{czachor97}, Lee and Chang-Young \cite{leechangyoung04}, and Ahn-Hwang-Lee-Moon \cite{ahnetal02,ahnetal03a,moonetal04} the violation of a Bell-inequality in the non-relativistic limit cannot generally be sustained in a relativistic setting if the same measurement directions are chosen. It is clear however, and has been discussed in Ref.~\cite{leechangyoung04}, that by appropriately rotating the measurement directions, the maximal violation can be recovered in all inertial frames. We want to explain now how this arises in our context of using the Wigner-rotations introduced in Section \ref{sec:wigner rotations}.\\

Let us consider the same particle as described in three different inertial frames $S$, $S^{\,\prime}$, and $S^{\,\prime\prime}$, and we begin with the \emph{rest frame $S$} of the particle, there the particle's momentum is given by
\begin{equation}
k\,=\,\begin{pmatrix} m \\ \vec{0} \end{pmatrix}
\label{eq:rest frame momentum}
\end{equation}
and the quantum state of the particle is
\begin{equation}
\left|\,\psi_{\,\mathrm{k}}\,\right\rangle\,=\,
\left|\,k\,\right\rangle\,\left|\,s\,\right\rangle \ \ \  .
\label{eq:rest frame state}
\end{equation}

Now suppose the spin of the particle is measured (since the particle is at rest one can think of the detector moving around the particle) along direction $\vec{a}$ (as seen by the particle), the correct observable would then be
\begin{equation}
\hat{a}\,=\,\frac{\vec{a}\cdot\vec{\sigma}}{|\vec{a}|} \ \ \ .
\label{eq:rest frame observable}
\end{equation}

Next we study the same situation from the \emph{frame $S^{\,\prime}$}, where the particle has momentum
\begin{equation}
p\,=\,L(p)\,k\,=\,\begin{pmatrix} \,p^{0} \\ \vec{p} \end{pmatrix} \ \ \ ,
\label{eq:frame S prime momentum}
\end{equation}
and the state is given by
\begin{equation}
\left|\,\psi_{\,\mathrm{p}}\,\right\rangle\,=\,U(L(p))\,\left|\,\psi_{\,\mathrm{k}}\,\right\rangle\,=\,
\left|\,p\,\right\rangle\,\left|\,s\,\right\rangle \ \ \ .
\label{eq:frame S prime state}
\end{equation}
No Wigner-rotation occurs in Eq.~(\ref{eq:frame S prime state}) since the Wigner-rotation angle vanishes for transformations from the rest frame to the moving frame. Since now the particle is moving, to measure along the same direction as the observer in $S$ we need to choose our measurement direction as $\vec{a}^{\,\prime}$, the spatial part of $a^{\,\prime}=L(p)a$. Clearly, by inserting it into the equation for the observable $\hat{a}^{\,\prime}$ in $S^{\,\prime}$ (the analog to (\ref{eq:pauli-lubanski-direction from rest frame})) we get
\begin{equation}
\hat{a}^{\,\prime}\,=\,\frac{(L^{-1}(\mathrm{p})a^{\,\prime})^{i}\sigma_{i}}
{|(L^{-1}(\mathrm{p})a^{\,\prime})^{j}|}\,=\,
\frac{a^{i}\sigma_{i}}{|a^{j}|}\,=\,\hat{a} \ \ \,.
\label{eq:frame S prime observable}
\end{equation}
Acknowledging the fact that the spin state remains unchanged we see that the expectation value of $\hat{a}^{\,\prime}$ in $S^{\,\prime}$ and $\hat{a}$ in $S$ provides the same result.\\

Let us finally proceed to study the situation from the third reference frame, $S^{\,\prime\prime}$, related to the frame $S^{\,\prime}$ by a Lorentz transformation $\Lambda$, such that the particle momentum in $S^{\,\prime\prime}$ is
\begin{equation}
p_{\Lambda}\,=\,\Lambda\,p\,=\,\Lambda\,L(p)\,k\,=
\,\begin{pmatrix} \,p_{\Lambda}^{0} \\ \vec{p}_{\Lambda} \end{pmatrix}\ \ \ .
\label{eq:frame S primeprime momentum}
\end{equation}
The state of the particle undergoes a Wigner-rotation $\textrm{W}(\Lambda ,p)$, i.e. recalling Eq.~(\ref{eq:unitary rep of lorentz trafo continued}) we find
\begin{equation}
\left|\,\psi_{\,\mathrm{p_{\Lambda}}}\,\right\rangle\,=\,
U(\Lambda)\,\left|\,\psi_{\,\mathrm{p}}\,\right\rangle\,=\,
U(L(p_{\Lambda}))\,U(\textrm{W}(\Lambda ,p))\,\left|\,\psi_{\,\mathrm{k}}\,\right\rangle\,=\,
\left|\,p_{\Lambda}\,\right\rangle\,
U(\textrm{W}(\Lambda ,p))\,\left|\,s\,\right\rangle \ \ \,,
\label{eq:frame S primeprime state}
\end{equation}
where $\textrm{W}(\Lambda ,p)=L^{-1}(\mathrm{p_{\Lambda}})\Lambda L(\mathrm{p})$. The observer in $S^{\,\prime\prime}$ will then see the measurement direction chosen by the observers in $S^{\,\prime}$ and the rest frame $S$ to be
\begin{equation}
a^{\,\prime\prime}\,=\,\Lambda\,a^{\,\prime}\,=\,\Lambda\,L(\mathrm{p})\,a \ \ \,.
\label{eq:frame S primeprime measurement direction}
\end{equation}
Since the particle has momentum $\mathrm{p_{\Lambda}}$ in $S^{\,\prime\prime}$, the corresponding spin observable for the chosen direction $a^{\,\prime\prime}$ according to equation (\ref{eq:pauli-lubanski-direction from rest frame}) is
\begin{equation}
\hat{a}^{\,\prime\prime}\,=\,\frac{(L^{-1}(\mathrm{p_{\Lambda}})a^{\,\prime\prime})^{i}\sigma_{i}}
{|(L^{-1}(\mathrm{p_{\Lambda}})a^{\,\prime\prime})^{j}|}\,=\,
\frac{(\textrm{W}(\Lambda ,p)a)^{i}\sigma_{i}}{|(\textrm{W}(\Lambda ,p)a)^{j}|} \ \ \,,
\label{eq:frame S primeprime observable}
\end{equation}
where we used the equations (\ref{eq:frame S primeprime measurement direction}) and (\ref{eq:wigner rotation}) in the second step. But $\textrm{W}(\Lambda ,p)$ is a rotation, therefore it does not change the norm of $\vec{a}$ and we may write
\begin{equation}
\hat{a}^{\,\prime\prime}\,=\,\left(R(\textrm{W}(\Lambda ,p))\frac{\vec{a}}{|\vec{a}|}\right)\cdot\vec{\sigma}
\;\,=\,\;U(\textrm{W}(\Lambda ,p))\,
\left[\frac{\vec{a}\cdot\vec{\sigma}}{|\vec{a}|}\right]\,
U^{\dagger}(\textrm{W}(\Lambda ,p)) \ \ \,.
\label{eq:frame S primeprime observable related to S}
\end{equation}

Eq.~(\ref{eq:frame S primeprime observable related to S}) is now the important relation between the observables, it implies that the expectation values of the observables $\hat{a}^{\,\prime\prime}$ in $S^{\,\prime\prime}$ and $\hat{a}$ in $S$ clearly coincide since the unitary transformations of the states in $S^{\,\prime\prime}$, depending now on the Wigner-rotations (as shown in Eq.~(\ref{eq:frame S primeprime state})), compensate precisely the corresponding ones in Eq.~(\ref{eq:frame S primeprime observable related to S}). It is crucial to emphasize here, that although the observable $\hat{a}^{\,\prime\prime}$ depends on the momentum of the particle, the measurement direction $a^{\,\prime\prime}$, corresponding to this observable, which is chosen by the observer in $S^{\,\prime\prime}$, does not depend on the momentum of the particle.\\

Finally, we have to consider the combined observables for tensor products of states like $\left|\,p_{+}\,\right\rangle\,\left|\,s_{+}\,\right\rangle\otimes\,
\left|\,p_{-}\,\right\rangle\,\left|\,s_{-}\,\right\rangle\,$, thus each observable acts separately in its subspace depending on the momentum, i.e.
\begin{eqnarray}
&&\left (\hat{a}(q)\otimes\hat{b}(q) \right ) \;\; \left|\,p_{+}\,\right\rangle\,\left|\,s_{+}\,\right\rangle\otimes\,
\left|\,p_{-}\,\right\rangle\,\left|\,s_{-}\,\right\rangle \,=\, \nonumber \\
&&\left|\,p_{+}\,\right\rangle\,
\frac{\vec{a}\cdot\vec{\sigma}_{p_{+}}}{|\lambda(\vec{a}\cdot\vec{\sigma}_{p_{+}})|}\,
\left|\,s_{+}\,\right\rangle \otimes
\left|\,p_{-}\,\right\rangle
\frac{\vec{b}\cdot\vec{\sigma}_{p_{-}}}{|\lambda(\vec{b}\cdot\vec{\sigma}_{p_{-}})|}\,
\left|\,s_{-}\,\right\rangle \ \ \ ,
\label{eq:action of tensorproduct normalized relativistic spin observable}
\end{eqnarray}
according to Eq.~(\ref{eq:normalized relativistic spin observable}). We obviously find that all spin measurements along a certain direction are independent of the choice of reference frame if the spin observable is given by (\ref{eq:normalized relativistic spin observable}) and (\ref{eq:pauli-lubanski-direction from rest frame}) and the measurement directions are transformed accordingly for the differently chosen frames. Most importantly, this implies that the maximal violation of the Bell-inequality is independent of the chosen frame and can always be recovered for the right choice of directions.\\

For example, considering again the scheme of the two observers Alice and Bob in the frame $S^{\,\prime}$ where particle beams are emitted along the ($\pm z$)-direction, and let us take the measurement directions $\vec{a}, \vec{\alpha}, \vec{b}, \vec{\beta}$ for the Bell-observable in the CHSH-inequality \cite{bell-book,chsh69}, then we have
\begin{equation}
S(\vec{a},\vec{\alpha},\vec{b},\vec{\beta})\,=\,
|\,E(\vec{a},\vec{b})\,-\,E(\vec{a},\vec{\beta})\,|\,+\,
|\,E(\vec{\alpha},\vec{\beta})\,+\,E(\vec{\alpha},\vec{b})\,| \,\leq\,2 \ \ \ .
\label{eq:CHSH inequality}
\end{equation}
We also choose the measurement directions in the $x-y$ plane, then the spin observable (\ref{eq:normalized relativistic spin observable}) reduces to the non-relativistic spin observable and there is no change in the directions, i.e.
\begin{equation}
a^{\prime}\,=\,a\,=\,
\begin{pmatrix} 0 \\ a_{\mathrm{x}} \\ a_{\mathrm{y}} \\ 0 \end{pmatrix}\ \ \ , \ \ \
b^{\prime}\,=\,b\,=\,
\begin{pmatrix} 0 \\ b_{\mathrm{x}} \\ b_{\mathrm{y}} \\ 0 \end{pmatrix}\ \ \ .
\label{eq:measurement directions in xy plane}
\end{equation}
Furthermore, assuming that the source, which is at rest in the frame of A and B, produces particles in the singlet spin state $|\,\psi^{-}\left.\right\rangle$ we get the familiar expectation value
\begin{equation}
E(\vec{a},\vec{b})\,=\,
\left\langle\right.\psi^{-}\,|\,\left\langle\right. k\,|\;\,\hat{a}\otimes\hat{b}\;\,
|\,k\,\left.\right\rangle\,|\,\psi^{-}\left.\right\rangle\,=\,
-\,\frac{\vec{a}\cdot\vec{b}}{|\vec{a}|\cdot|\vec{b}|}\ \ \ ,
\label{eq:psi minus exp value nonrelativistic}
\end{equation}
such that A and B will be able to violate the CHSH inequality maximally by $2\sqrt{2}$ for suitable measurement directions in that plane, regardless which exact parameter is chosen in the momentum state of (\ref{eq:initial momentum state}).\\

Now, let us regard the situation where A and B are moving in the $x$-direction w.r.t. the source, the state observed by A and B gets Wigner rotated according to Eq.~(\ref{eq:boosted general total state}) (and explicitly given in Eq.~(\ref{eq:boosted bell type spin total state}))
\begin{eqnarray}
\left|\,\psi^{\Lambda}\,\right\rangle &=&
        \cos\alpha\left|\,\Lambda p_{+},\Lambda p_{-}\,\right\rangle
		U_{+}\otimes U_{-}\left|\,\psi^{\,-}\right\rangle\,+
        \,\sin\alpha\left|\,\Lambda p_{-},\Lambda p_{+}\,\right\rangle
		U_{-}\otimes U_{+}\left|\,\psi^{\,-}\right\rangle\ \ \ .
\label{eq:example transformed spin state psi minus}
\end{eqnarray}
Then the expectation value in the moving system $S^{\,\prime\prime}$ coincides with the one in $S^{\,\prime}$ and $S$
\begin{equation}
E^{\,\prime\prime}(\vec{a}^{\,\prime\prime},\vec{b}^{\,\prime\prime})\,=\,
\left\langle\right.\psi^{\Lambda}\,|\;\,\hat{a}^{\,\prime\prime}\otimes\hat{b}^{\,\prime\prime}\;\,
|\,\psi^{\Lambda}\,\left.\right\rangle\,=\,
-\,\frac{\vec{a}\cdot\vec{b}}{|\vec{a}|\cdot|\vec{b}|}\,=\,E(\vec{a},\vec{b}) \ \ \ ,
\label{eq:psi minus exp value in relativistic system}
\end{equation}
which can be easily seen by using Eqs.~(\ref{eq:frame S primeprime state}), (\ref{eq:frame S primeprime observable related to S}) and (\ref{eq:action of tensorproduct normalized relativistic spin observable}), i.e. the unitary transformations in the states, Eq.~(\ref{eq:frame S primeprime state}), just compensate the corresponding ones in the observable, Eq.~(\ref{eq:frame S primeprime observable related to S}).\\

Of course, would the moving observers perform their measurements in directions given by the components of $\vec{a},\vec{\alpha},\vec{b},\vec{\beta}$ as seen in the rest frame of the source, the CHSH-inequality could not generally be maximally violated. When restricting ourselves to measurements in the $x-y$ plane, the directions chosen by the observer in $S^{\,\prime\prime}$ are related to the ones in $S^{\,\prime}$ by the Lorentz transformation $\Lambda$
\begin{equation}
a^{\,\prime\prime}\,=\,\Lambda a^{\,\prime}\,= \,
\begin{pmatrix} \cosh\xi    & -\sinh\xi  & 0 & 0 \\
                -\sinh\xi & \cosh\xi     & 0 & 0 \\
                0           & 0            & 1 & 0 \\
                0           & 0            & 0 & 1 \\
\end{pmatrix}
\begin{pmatrix} 0 \\ a_{\mathrm{x}} \\ a_{\mathrm{y}} \\ 0 \end{pmatrix}\,=\,
\begin{pmatrix} -\sinh\xi\,a_{\mathrm{x}} \\ \ \cosh\xi\,a_{\mathrm{x}} \\ \ \ a_{\mathrm{y}} \\ 0 \end{pmatrix}
\ \ \ ,
\label{eq:transformed xy plane measurement directions}
\end{equation}
which is a boost to a frame moving along the ($+x$)-direction with rapidity $\xi$. Thus the measurement direction in the moving system $S^{\,\prime\prime}$ with a suitable normalization is given by
\begin{equation}
\frac{\vec{a}^{\,\prime\prime}}{|\vec{a}^{\,\prime\prime}|}\,=\,
\frac{1}{\sqrt{(\cosh\xi\,a_{\mathrm{x}})^{2}+a_{\mathrm{y}}^{2}}}
\begin{pmatrix} \cosh\xi\,a_{\mathrm{x}} \\ \ \ a_{\mathrm{y}} \\ 0 \end{pmatrix} \ \ \ .
\label{eq:corrected xy plane measurement directions}
\end{equation}
One can see here, that directions chosen purely in the $x$ or $y$ axis respectively are not changing, while all other directions in the $x-y$ plane are tilted towards the $x$-axis.

\section{Conclusion}
\label{conclusion}

The entanglement of a two particle state, consisting of spin--$\tfrac{1}{2}$ particles, cannot generally be considered to be Lorentz-invariant, although certain states and certain partitions retain their entanglement. In this paper we have fully worked out the dependence of the entanglement change on a number of parameters, including the explicit form of the spin- and momentum-state, the choice of reference frame, as well as the partitions of the 4 qubits considered. The connection of the entanglement change to the invariant maximal violation of a Bell inequality also becomes immediately clear, since we also demonstrated that the entanglement between the partitions correlated with the Bell inequality violation remains unchanged.\\

In particular, we have computed that the overall change in entanglement is nonzero generally and only the entanglement between certain partitions of the total Hilbert space, i.e. the Alice-Bob partition and the partition into (spin-A+mom-B) / (mom-A+spin-B) subspaces, remains invariant. There is an entanglement change, however, in the other partitions, in the four (1 qubit) / (3 qubits) partitions and in the (2 spins) / (2 momenta) partition. Interestingly, if the Lorentz boosts for observer and particle reach the speed of light, i.e. $\delta=\frac{\pi}{2}$, the entanglement changes of both partitions agree, such that for particles moving at the speed of light the entanglement change due to the shifted reference frame can be traced back to a change in the entanglement between the spins and the momenta.\\

The invariance of the Bell inequality violation in different reference frames is achieved by Lorentz transforming both the states \emph{and} the observables such that each observer will measure the same expectation values if the correct measurement directions are chosen. However, since the entanglement of the reduced spin density matrix, i.e. here represented by the entanglement of the (1 qubit) / (3 qubits) partition, changes (see also \cite{gingrichadami02}), it seems incorrect to assume, that the Bell inequality is sensitive to the entanglement of the spin state alone, but rather that the entanglement between the particles of Alice and Bob corresponds to the violation of the Bell inequality.\\

But nonetheless the entanglement of the spins is playing a crucial role in violating Bell inequalities. Consider e.g. an initial state, composed of a maximally entangled momentum state, i.e. $\alpha=\tfrac{\pi}{4}$, and a separable spin state, e.g. if $\beta=0$. Since the Bell inequality would not be violated for any combination of measurement directions in the initial frame, it will not be violated in any other frame, regardless of the fact that the entanglement of some partitions might change or that the entanglement with respect to the Alice-Bob partition is non-zero. Invoking the theorem of Gingrich and Adami \cite{gingrichadami02}, which states that the entanglement between spins and momenta must be non-zero in order for the spin entanglement to increase under Lorentz transformations, we see that the entanglement change cannot increase the spin entanglement in this situation. We therefore conclude that although the violation of Bell inequalities depends on the overall entanglement between the two particles, it cannot be brought about by momentum entanglement alone.\\

As mentioned before, the Bell inequality is only sensitive to the invariant entanglement between Alice and Bob. However, for the states of Section \ref{subsec:bell type states} the entanglement-egg-tray (see Fig.~\ref{fig:entanglementchangebelltype}) suggests that the entanglement of other partitions changes, but it does so only if there is some entanglement in the momentum state initially, and only, if the spin entanglement is not maximal to begin with. This is the reason why the initial states with no momentum entanglement, i.e. $\alpha=\frac{n\pi}{2}$, and those with maximal spin entanglement, i.e. $\beta=\frac{(2n+1)\pi}{4}$, show no change at all in entanglement. For all the other combinations of $\alpha$ and $\beta$ there is an increase in entanglement between the four qubits, in particular if the spin state is separable initially. Therefore it would be of interest whether another type of inequality, e.g. an entanglement witness inequality \cite{bertlmann-narnhofer-thirring02,bruss02,terhal00,horodecki-mpr96}, could be found, which is sensitive to the entanglement between spins and momenta. The such found entanglement witnesses, which might correspond to the observables in an experiment, could detect the discussed change in the entanglement of the particular partitions of the qubits.\\

It remains to note that the initially chosen momentum state, crucial to the entanglement change, requires the particles to be distinguishable. Although this somehow restricts the choice of physical systems for which such a change is possible, the distinctive property of the two particles could be some additional quantum number, invariant under Lorentz transformations. It remains to be seen whether similar effects emerge by considering different modes of distinguishable particles in a second quantization formalism.\\

\begin{acknowledgments}

We particularly want to thank Jakob Yngvason, Ivette Fuentes-Schuller, and Frank Verstraete for insightful discussions, helpful remarks and encouragement for this work. Marcus Huber gratefully acknowledges the
Austrian Fund project FWF-P21947N16. Beatrix C. Hiesmayr gratefully acknowledges the fellowship MOEL 428.

\end{acknowledgments}

\end{document}